\documentclass[mathleft]{an}
\usepackage{graphicx}
\usepackage{times}
\usepackage{natbib}
\bibpunct{(}{)}{;}{a}{}{}

\begin{document}

\Pagespan{1}{}
\Yearpublication{2010}%
\Yearsubmission{2009}%
\Month{11}%
\Volume{}%
\Issue{}%

\title{Expanding atmosphere models for SSS spectra of novae}

\author{D.R. van Rossum\inst{1}\fnmsep\thanks{Corresponding author: drossum@hs.uni-hamburg.de} \& J.-U. Ness\inst{2}}
\titlerunning{Expanding atmosphere models for SSS spectra of novae}
\authorrunning{D.R. van Rossum}
\institute{Hamburger Sternwarte, Gojenbergsweg 112, 21029 Hamburg, Germany
\and
European Space Astronomy Centre,
P.O. Box 78, 28691 Villanueva de la Ca\~nada, Madrid, Spain}

\received{\today}
\accepted{11 Nov 2005}
\publonline{later}


\abstract{
Super Soft Source (SSS) spectra are powered by nuclear burning on the surface
of a white dwarf. The released energy causes a radiatively-driven
wind that leads to a radially extended atmosphere around the white dwarf.
Significant blue shifts in photospheric absorption lines are found in the spectra of novae during their SSS phase, being an evidence of continued mass loss in this phase.
We present spherically symmetric {\tt PHOENIX} models that account for the expansion of the ejecta.
A comparison to a plane parallel, hydrostatic atmosphere model demonstrates that the mass loss can have a significant impact on the model spectra.
The dynamic model yields less pronounced absorption edges, and harder X-ray spectra are the result.
Therefore, lower effective temperatures are needed to explain the observed spectra.
Although both types of models are yet to be fine-tuned in order to accurately determine best fit parameters, the implications on the chemical abundances are going in opposite directions.
With the expanding models the requirement for strong depletion of the crucial elements that cause these edges is now avoidable.
}

\maketitle

\section{Introduction}

SSS are commonly accepted to be
supported by nuclear fusion on the surface of a white dwarf, where
the hydrogen-rich burning material is provided by accretion from a companion star (Kahabka \& van den Heuvel 1997).
Observed SSS spectra show the direct signatures of nuclear burning (Brems\-strahlung spectra of $\sim$ $10-30$\,keV).
The white dwarf is surrounded by an optically thick atmosphere that emits in soft X-rays.
In addition to the so-called persistent SSSs like Cal\,83, some
classical novae were observed to go through a phase of
SSS emission, when their spectrum resembles that of Cal\,83.
This is not surprising, as it is expected that the opacity of
the ejecta is changing with time, which leads to a
shift of the peak of the spectral energy distribution from
the optical via the UV to the soft X-rays (see e.g.,
\citealt{gallstar78}). The X-ray emitting phase is reached when
the radius of the photosphere has receded into sufficiently hot
layers. This is a consequence of declining mass-loss rate which leads
to a decrease of the density in the envelope. Novae have been observed
with extremely bright SSS emission, e.g. V1974\,Cyg by (Krautter et
al. (1996), much brighter than the persistent SSSs. The
brightness of nova SSS spectra makes them ideal
targets for spectral analysis. In this paper we
focus on the novae RS\,Oph and V2491\,Cyg.


The X-ray gratings on board XMM-Newton and Chandra yield a spectral resolution that is two orders of magnitude higher than precursor X-ray satellites ROSAT and BeppoSax.
The first X-ray grating spectrum of a nova in outburst, presented by \cite{v4743}, yields deep absorption lines.
These high quality spectra are oftentimes interpreted with the help of hydrostatic atmosphere models, e.g. \citep{rauch05} and \citep{nelson07}.
But even if, in some cases, atmosphere models yield better representation of the data than blackbody fits,
a good fit alone is not a sufficient justification for the model to be realistic.
%

An important achievement from the grating spectrometers is the detection of line shifts in the SSS spectra of novae, as first seen by \cite{v4743} and illustrated by \cite{ness_rsoph} and \cite{ness_SSS}.
In hydrostatic models these line shifts are not recovered.
Blue-shifting the whole synthetic spectrum with a fixed amount in order to better fit the observation does not physically represent an expanding atmosphere.
The expansion is a spherical effect in which some parts of the atmosphere move towards the observer while some other parts move away.
Furthermore, the expansion affects the radial extension of the atmosphere.
In this paper we present expanding atmosphere models, compare
them to static models, and show that the results obtained
from fits to observations are significantly different.

\section{Expanding atmosphere models}

We present first results of new models that were computed with
the general-purpose, state-of-the-art stellar atmosphere code
{\tt PHOENIX} (Hauschildt \& Baron 1992, Hauschildt \& Baron
1999, Hauschildt \& Baron 2004).
The radiation transfer
equations are solved using the operator splitting method.
Spherical symmetry is assumed for the atmosphere, and NLTE computation is based on the assumption of statistical equilibrium.
{\tt PHOENIX} was used to model X-ray novae before \citep{petz05}.

Since 2005, the code has been improved significantly, namely, NLTE has been reimplemented with new
rates and opacities, a new temperature correction
method was used, broad lines are handled differently,
and a new (hybrid, see below) atmosphere construction method has been
implemented, based on the idea of \cite{Aufdenberg00}. The improved code is physically more
realistic and is a factor 15-45 faster. For details
we refer to van Rossum (Ph.D. thesis in preparation).

The hybrid atmosphere consists of a hydrostatic base with an expanding envelope on top.
The hydrostatics are parametrised in the classic way by
$\log(g)$, $r_0$ and $T_{\rm eff}$.
Here, $r_0$ is the outer radius of the static core. For the expanding
envelope, a velocity field is defined, and from the continuity equation for a constant mass loss rate, $\dot{M}$, a density profile follows.
The velocity fields assumed in the models presented here are described by the beta-law (Lamers \& Cassinelli 1999).
\begin{equation}
 v(r\ge r_0)=v_\infty(1-r_0/r)^\beta\,.
\end{equation}
The additional parameters that describe the expanding envelope are the terminal velocity $v_\infty$ and the exponent $\beta$.

 In this paper we present the first results of the models, based on which
more refined models can be calculated. As a start, we assumed solar abundances
of the elements H, He, C, N, O, Ne, Al, Mg, Si, S, Ar, Ca,
and Fe (yielding roughly 7500 independent levels with 150,000 transitions,
all calculated in NLTE).
We emphasise that the models presented in this paper have not been fine tuned, i.e., the full parameter space has not been explored yet.

\section{Results}
\begin{figure}[!ht]
\resizebox{\hsize}{!}{\rotatebox{90}{\includegraphics{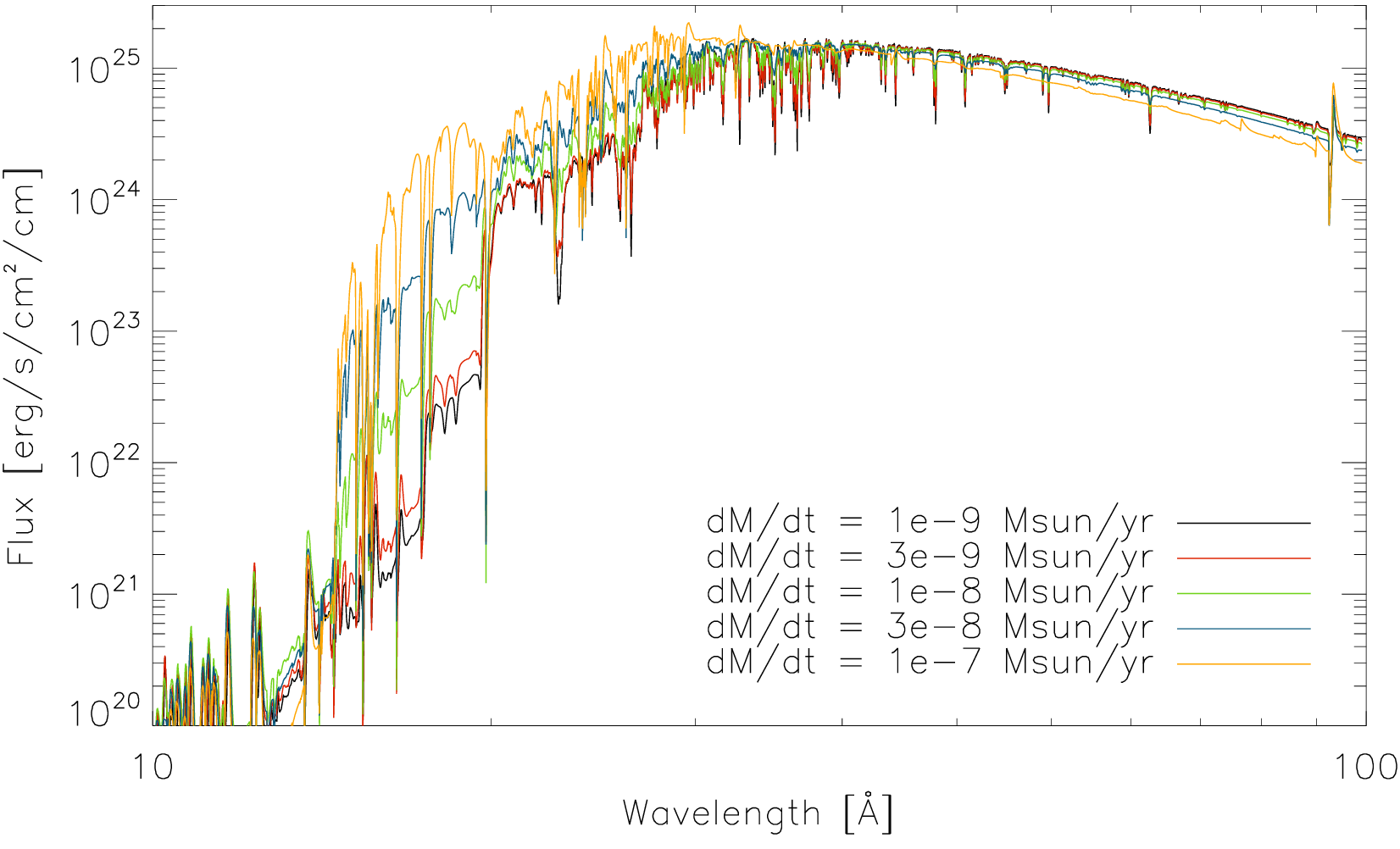}}}
\caption{\label{dm}
Example of the impact of the mass-loss rate on the model spectrum.
A range of models with $T_{\rm eff}=6.0\cdot 10^5$\,K and $v_\infty = 2400$ km/s and different mass-loss rates is shown.
With increasing mass-loss rate the ionisation edges at 25.3\,\AA{}, 18.6\,\AA{} and 16.8\,\AA{} (from the elements C, N and O, respectively) become weaker, and the flux between 14 and 30\AA{} increases.
}
\end{figure}

 In Fig.~\ref{dm}, we illustrate the effect of the mass loss parameter on the model spectrum.
By increasing the mass-loss rate the ionisation edges become weaker and thus the spectra become harder.
As a consequence, dynamic models that are fitted to observed spectra typically yield a lower effective temperature than static models.


\begin{figure*}[!t]
\resizebox{\hsize}{!}{\rotatebox{90}{\includegraphics{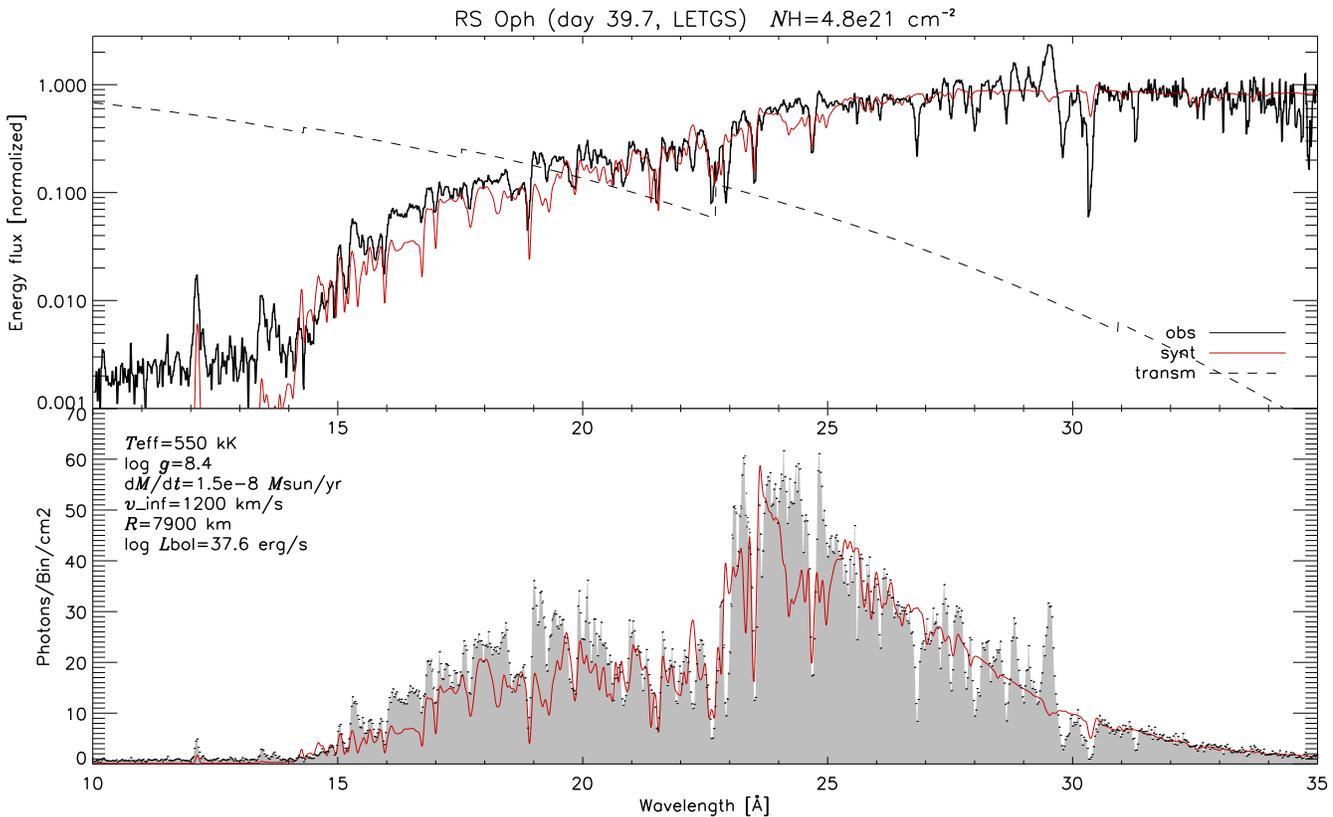}}}
\caption{\label{RSOph}Comparison of a first cut {\tt PHOENIX} model spectrum with the observation of RS\,Oph. 
The binwidth is 0.025\,\AA{}.
In the top panel, the observed spectrum is corrected for
interstellar absorption, yielding the non-absorbed spectrum.
The interstellar transmission coefficient 
is shown as dashed line.
In the bottom panel, the model was corrected for absorption.
The continuum slope is reproduced by the model longwards of 14\,\AA{}.
At shorter wavelengths the flux is dominated by sources other than the SSS atmosphere, e.g. emission from the shock.
These contributions are observed already before the SSS phase starts and are not included in the model.
}
\end{figure*}

\begin{figure*}[!t]
\resizebox{\hsize}{!}{\rotatebox{90}{\includegraphics{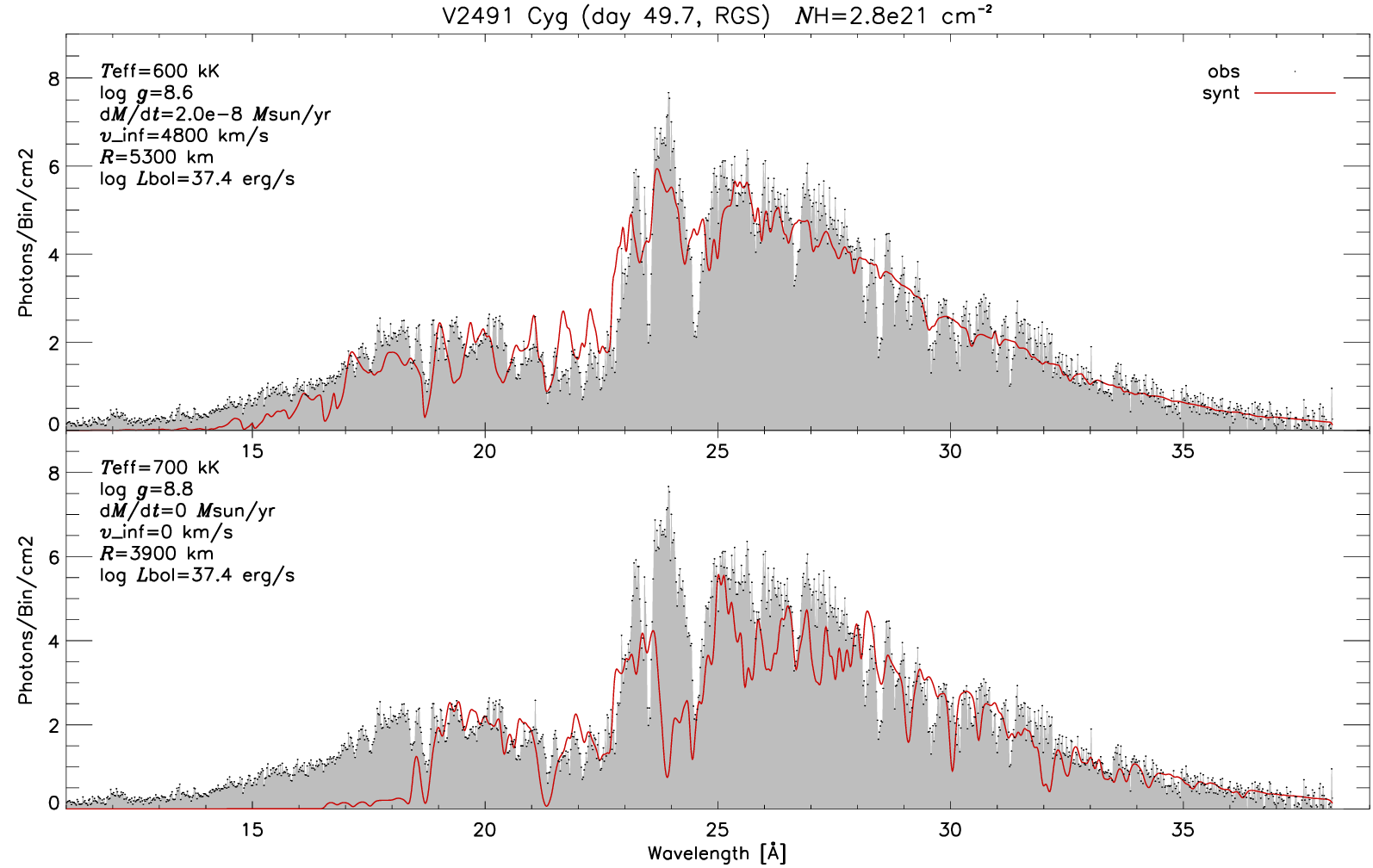}}}
\caption{\label{V2491}XMM-Newton RGS spectrum of V2491\,Cyg
(2008). 
The binwidth is 0.02\,\AA{}. 
Two different models
are shown (red curves), a radially extended model with mass loss (top
panel) and a white dwarf (compact, hydrostatic) model (bottom
panel). No good fit could be found with white dwarf models.
The best temperature is much higher and the radius much smaller than for the model with mass loss.
One important difference is the absence of \emph{strong} ionisation edges in the dynamical model, like at 25.3\,\AA{}, 18.6\,\AA{} and 16.8\,\AA{}.
Note that the models use identical (solar) abundances but give very different implications on being either under- or overabundant (see the discussion in the text).
}
\end{figure*}

In Fig.~\ref{RSOph}, we
show the observation of RS\,Oph that was taken on
day 39.7 after outburst (for an observation log we refer to
Ness 2009 and references therein). A \emph{small} grid of models was
calculated with a terminal velocity of $v_\infty = 1200$ km/s, according to the blue-shifts observed for this object (\citealt{ness_rsoph}).
We achieve the best agreement with
$T_{\rm eff}=5.5\cdot 10^5$\,K, $\log g=8.4$, a wind mass loss rate of
$\dot{M}=1.5\cdot 10^{-8}$\,M$_\odot/$yr,
$R_{\rm eff}=7900$\,km,
and $N_{\rm H}=4.8\cdot 10^{21}$\,cm$^{-2}$.

 Our interstellar extinction model is based on absorption cross sections in the X-ray domain compiled by
Balucinska-Church \& McCammon (1992)
for 17 astrophysically important species.
Specifically, the effective extinction curve was calculated with Balu\-cins\-ka-Church \& McCammon's code, assuming solar abundances.
The total extinction is then proportional to the hydrogen column density.
We would like to point out that, owing to the large
interstellar column density, the soft part of the spectrum
is significantly underrepresented when fitting a model to
an absorbed spectrum. In the top panel of Fig.~\ref{RSOph},
it can be seen that the emission that is produced by the
source (thus before traversing through the interstellar
medium), peaks at much longer wavelengths than the observed
spectrum (bottom panel in grey shadings;
see also Fig. 1 in Ness 2009). Since the ultimate
goal is to reproduce the original (unabsorbed) spectrum (note that the interstellar extinction has no physical relation to the object we try to model),
reproduction of the soft tail is at least as important as the rest of the spectrum.
Since the soft tail is primarily determined by the interstellar extinction, this leads to the conclusion that $N_{\rm H}$ cannot be treated as a free parameter in the fitting procedure.
Instead, $N_{\rm H}$ must be determined to reproduce the soft tail.
This restriction in free parameter space grievely limits the possibilities in finding a good fit.
Accounting for this effect, especially, hydrostatic models turn out to poorly fit the observations (see for example the bottom panel in Fig. \ref{V2491}).
The usual goodness criteria (e.g., $\chi^2$), determined from fitting a model to the absorbed spectrum \emph{only} yield a description of how well the data, especially the brightest parts of the spectrum, are reproduced.
But that does \emph{not} necessarily indicate whether the physical description of the object is realistic or not.
We therefore fit the parameter $N_{\rm H}$ by eye, considering both the absorbed \emph{and} unabsorbed fit quality.

 In Fig.~\ref{V2491}, the results for the more recent nova
V2491\,Cyg (2008) are presented, where we compare with an
XMM-Newton observation taken on day 49.7 after outburst
(grey shadings). In the top panel, the best model that has
so far been found from a \emph{small} grid of models with $v_\infty = 4800$ km/s is shown, yielding the parameters
$T_{\rm eff}=6.0\cdot 10^5$\,K, $\log g=8.6$, a wind mass loss rate of
$\dot{M}=2\cdot 10^{-8} \mathrm{ M}_\odot/$yr,
$R_{\rm eff}=5300$\,km,
and $N_{\rm H}=2.8\cdot 10^{21}$\,cm$^{-2}$.
Good agreement between observation and model is found, and it can be expected to yet improve significantly when non-solar abundances are used and the atmospheric structure is fine-tuned to the observation.
In the actual comparison in Fig. \ref{V2491}, many lines do not yet accurately match their observed strengths and the continuum is too faint for shorter wavelengths.
The latter point may, apart from the model that lacks tuning, be attributed to non-atmospheric X-ray sources that are not included in the model.
In the bottom panel, the same observation is shown in comparison to a plane parallel and static {\tt PHOENIX} model with no mass loss that has been tuned to a similar degree as the expanding model.
The temperature in the static model is significantly higher, which is necessary to compensate the lack of hard emission shortwards of the N\,{\sc vii} and O\,{\sc vii} absorption edges at 18.6\,\AA\ and 16.8\,\AA, respectively.
As a consequence, a smaller effective radius is needed.
Note that the obtained bolometric luminosities are the same, which is primarily a result of the constant $N_{\rm H}$ obtained for both fits.

But more interesting are the different implications for abundances.
For example, in the hydrostatic model the N\,{\sc vii} edge at 18.6\,\AA{} is too strong, and also the N\,{\sc vii} line at 24.8\,\AA{} and the N\,{\sc vi} line at 28.8\,\AA{}, leading to the interpretation that the N abundance must be subsolar.
In contrast, in the expanding model the N\,{\sc vii} edge at 18.6\,\AA{} is very weak, the N\,{\sc vi} edge at 22.5\,\AA{} is too weak and also the N\,{\sc vii} line at 24.8\,\AA{} and the N\,{\sc vi} line at 28.8\,\AA{}, so that according to the new type of models the N abundance should become supersolar.
These discrepancies must be examined in detail with the help of fine-tuned models, which are not yet available at the time of writing.

\vspace{-.5cm}
\section{Discussion and Conclusions}

The comparison between the hydrostatic and the new dynamic model in Fig.~\ref{V2491} demonstrates the essence of treating the mass-loss in the model atmosphere.
Obviously, the model spectra are sensitive to the details of the atmospheric structure.
Comparing observations to models that account for the mass-loss can lead to very different conclusions than comparison with static atmospheres.
Since the continued mass-loss is undeniably confirmed by the observations, this calls for caution with the interpretation of nova X-ray spectra using static atmospheres.
For example,
the conclusion from fine tuning of a static atmosphere model to observations of RS Oph, presented by \cite{nelson07}, that the C/N abundance ratio is
1000 times subsolar can expected to be very different when dynamical models are used.

The models presented in this paper all use solar abundances.
In the SSS phase the observed radiation originates from much deeper regions than in earlier stages after the outburst.
The material in those regions is a mixture of dred\-ged-up material from the white dwarf, accreted material from the companion, and reaccreted ejecta of the initial outburst.
Therefore, the chemical composition that is ``seen'' at this stage could be significantly different from other stages.
Pre\-sent\-ly, there are no detailed theoretical and observational constraints on this mixture.
Although solar abundances mi\-ght be unrealistic, it is still a good standard set to start with.
Detailed abundance analysis with both static and dynamic models will be needed in order to draw definitive conclusions.
This must be performed with simultaneous fine-tu\-ning of the atmospheric structure, a task that is scheduled for the near future.

 The preliminary models presented in this paper are a
promising start. More sophisticated and better fine-tuned
models will increase the accuracy of the
parameters and the reproduction of the observed spectra.



\small

\vspace{-.2cm}

\begin{thebibliography}{13}
\expandafter\ifx\csname natexlab\endcsname\relax\def\natexlab#1{#1}\fi

\vspace{-.4cm}

\bibitem[{{Aufdenberg}(2000)}]{Aufdenberg00}
{Aufdenberg}, J.P. 2000, Ph.D.~Thesis, Arizona State University


\bibitem[{{Balucinska-Church} \& {McCammon}(1992)}]{Church92}
{Balucinska-Church}, M., {McCammon},D. 1992, ApJ, 400, 699

\bibitem[{{Gallagher} \& {Starrfield}(1978)}]{gallstar78}
{Gallagher}, J.S., \& {Starrfield}, S. 1978, ARAA, 16, 171

\bibitem[{{Hartmann} \& {Heise}(1997)}]{hartheis97}
{Hartmann}, H.W., \& {Heise}, J. 1997, A\&A, 322, 591

\bibitem[{{Hauschildt} \& {Baron}(1992)}]{phx92}
{Hauschildt}, P.H., \& {Baron}, E. 1992, JQSRT, 47, 433

\bibitem[{{Hauschildt} \& {Baron}(1999)}]{phx_expand}
{Hauschildt}, P.H., \& {Baron}, E. 1999, Journal of Computational and Applied
  Mathematics, 109, 41

\bibitem[{{Hauschildt} \& {Baron}(2004)}]{phx04}
{Hauschildt}, P.H., \& {Baron}, E. 2004, A\&A, 417, 317

\bibitem[{{Kahabka} \& {van den Heuvel}(1997)}]{kahab}
{Kahabka}, P., \& {van den Heuvel}, E.P.J. 1997, ARA\&A 35, 69

\bibitem[{{Krautter} {et~al.}(1996)}]{krautt96}
{Krautter}, J., {\"Ogelman}, H., {Starrfield}, S., {Wichmann}, R., {Pfeffermann}, E. 1996, ApJ, 456, 788

\bibitem[{{Lamers} \& {Cassinelli}(1999)}]{Lamers99}
{Lamers}, H.J.G.L.M, \& {Cassinelli}, J.P. 1999, Freeman (San Francisco)

\bibitem[{{Nelson} {et~al.}(2008)}]{nelson07}
{Nelson}, T., {Orio}, M., {Cassinelli}, J. P., {Still}, M., {Leibowitz}, E., {Mucciarelli}, P. 2008, ApJ, 673, 1067

\bibitem[{{Ness}(2009)}]{ness_SSS}
{Ness}, J.-U. 2009, AN

\bibitem[{{Ness} {et~al.}(2007)}]{ness_rsoph}
{Ness}, J.-U., {Starrfield}, S., {Beardmore}, A. P., et al. 2007, ApJ, 665, 1334

\bibitem[{{Ness} {et~al.}(2003)}]{v4743}
{Ness}, J.-U., {Starrfield}, S., {Burwitz}, V., et al. 2003, ApJL, 594, L127

\bibitem[{{Petz} {et~al.}(2005)}]{petz05}
{Petz}, A. and {Hauschildt}, P.~H. and {Ness}, J.-U. and {Starrfield}, S. 2005, A\&A, 431, 321

\bibitem[{{Rauch} {et~al.}(2005)}]{rauch05}
{Rauch}, T., {Werner}, K., \& {Orio}, M. 2005, in AIP Conf. Proc. 774: X-ray
  Diagnostics of Astrophysical Plasmas: Theory, Experiment, and Observation,
  ed. R.~{Smith}, 361

\end{thebibliography}

\end{document}